\documentclass[aps,pra,reprint,superscriptaddress]{revtex4-1} %aps,prl,reprint
\usepackage{graphicx}
\usepackage[colorlinks,citecolor=blue,urlcolor=blue,linkcolor=blue]{hyperref}
\usepackage{siunitx}
\usepackage{amsmath}
\usepackage{braket}
\usepackage{booktabs}

\begin{document}

\title{Modally Resolved Fabry-Perot Experiment with Semiconductor Waveguides}

\author{B. Pressl}
\email{Benedikt.Pressl@uibk.ac.at}
\affiliation{Institut f\"ur Experimentalphysik, Universit\"at Innsbruck, Technikerstra\ss e 25, 6020 Innsbruck, Austria}

\author{T. G\"unthner}
\affiliation{Institut f\"ur Experimentalphysik, Universit\"at Innsbruck, Technikerstra\ss e 25, 6020 Innsbruck, Austria}

\author{K. Laiho}
\affiliation{Institut f\"ur Experimentalphysik, Universit\"at Innsbruck, Technikerstra\ss e 25, 6020 Innsbruck, Austria}

\author{J. Ge\ss ler}
\affiliation{Technische Physik, Universit\"at W\"urzburg, Am Hubland,  97074 W\"urzburg, Germany}

\author{M. Kamp}
\affiliation{Technische Physik, Universit\"at W\"urzburg, Am Hubland,  97074 W\"urzburg, Germany}

\author{S. H\"ofling}
\affiliation{Technische Physik, Universit\"at W\"urzburg, Am Hubland,  97074 W\"urzburg, Germany}
\affiliation{School of Physics $\&$ Astronomy, University of St Andrews, St Andrews, KY16 9SS, United~Kingdom}

\author{C. Schneider}
\affiliation{Technische Physik, Universit\"at W\"urzburg, Am Hubland,  97074 W\"urzburg, Germany}

\author{G. Weihs}
\affiliation{Institut f\"ur Experimentalphysik, Universit\"at Innsbruck, Technikerstra\ss e 25, 6020 Innsbruck, Austria}

%\author[1,*]{B. Pressl}
%\author[1]{T. G\"unthner}
%\author[1]{K. Laiho}
%\author[2]{J. Ge{\ss}ler}
%\author[2]{M. Kamp}
%\author[2,3]{S. H\"ofling}
%\author[2]{C. Schneider}
%\author[1]{G. Weihs}
%
%\affil[1]{Institut f\"ur Experimentalphysik, Universit\"at Innsbruck, Technikerstra{\ss}e 25, 6020 Innsbruck, Austria}
%\affil[2]{Technische Physik, Universit\"at W\"urzburg, Am Hubland, 97074 W\"urzburg, Germany}
%\affil[3]{School of Physics \& Astronomy, University of St Andrews, St Andrews, KY16 9SS, United Kingdom}
%\affil[*]{Corresponding author: Benedikt.Pressl@uibk.ac.at}

%\dates{Compiled \today}

%\ociscodes{(070.4790) Spectrum analysis; (130.3120) Integrated optics devices; (130.5990) Semiconductors; (130.2790) Guided waves; (230.7370) Waveguides}

%\doi{\url{http://dx.doi.org/10.1364/optica.XX.XXXXXX}}

% Abstract and Cover Letter guide line: https://www.osapublishing.org/optica/journal/optica/about.cfm#letter

% "In addition, the abstract should be accessible to a general audience. We suggest beginning by giving the context and purpose of the work before outlining the methodology and results. The abstract should conclude with a statement evaluating and indicating any implications of your result. The cover letter and manuscript abstract will be evaluated by at least two editorial board members to determine if the paper is approved for external peer review."

\begin{abstract}
Based on the interaction between different spatial modes, semiconductor Bragg-reflection waveguides provide a highly functional platform for non-linear optics. Therefore, the control and engineering of the properties of each spatial mode is essential. Despite the multimodeness of our waveguide, the well-established Fabry-Perot technique for recording fringes in the optical transmission spectrum can successfully be employed for a detailed linear optical characterization when combined with Fourier analysis. A prerequisite for the modal sensitivity  is a finely resolved transmission spectrum that is recorded over a broad frequency band.
Our results highlight how the features of different spatial modes, such as their loss characteristics and dispersion properties, can be separated from each other allowing their comparison.
The mode-resolved measurements are important for optimizing the performance of such multimode waveguides by tailoring the properties of their spatial modes.
\end{abstract}

%\setboolean{displaycopyright}{true}

%\begin{document}

\maketitle
%\thispagestyle{fancy}
%\ifthenelse{\boolean{shortarticle}}{\abscontent}{}

Waveguided light sources are widely employed in many applications both in the field of classical and quantum optics, since the light inside them can simultaneously be guided very flexibly and controlled strictly \cite{Stegeman1989, Anderson1995,D.K.Serkland1995, Sohler2008, Lederer2008}. The confined waveguide size results in a finite selection of transversal modes that can propagate through the structure, further determining whether only a single optical mode is supported or if a multitude of them is accepted.

The linear optical waveguide characteristics such as refractive index and its dispersion, transmission losses or the reflectivity at the end facets are relevant to practical purposes and important feedback for manufacturing.  One often employed loss measurement technique is the Fabry-Perot characterization that is suitable for structures with  well-cleaved, parallel end facets that have at least modest reflectivity \cite{Kaminow1978}. After travelling several round trips inside the waveguide light interferes, and fringes can be observed in the transmission if the optical length the light sees inside the resonator is changed for example by heating it \cite{Regener1985} or by altering the wavelength of light \cite{SarehTaebi2008}. The loss characteristics can then be extracted from the fringe contrast. In the latter case this information can also be extracted via the Fourier transformation of the transmission spectrum \cite{Hofstetter1997}, where peaks occur at integer multiples of the resonator's optical length. Furthermore, the ratio of the heights of subsequent peaks is connected to the loss characteristics. With this technique, properties of sophisticated structures such as single-mode semiconductor lasers, surface plasmon resonators and ridge waveguides have been investigated and indications of cavity defects have been observed \cite{Hofstetter1998, Chen2005, Allione2008, Bijlani2009}.

Regarding waveguides, their typically very rich spatial mode structure \cite{Anderson1971, Roelofs1994, Banaszek2001} may pose a significant challenge in their application. For single mode waveguides a clear fringe contrast can be recorded in the Fabry-Perot transmission test. However, for multimode waveguides this is not the case \cite{Rossi2005, Abolghasem2009}.
Nevertheless, the mode-resolved characteristics are of great interest, especially for non-linear optical waveguides that are utilized as frequency-converted photon sources. Thus, the characteristics of all interacting modes have to be carefully studied for optimizing the conversion processes \cite{M.Karpinski2009, Christ2009, Mosley2009, Machulka2013}. Despite sophisticated fabrication techniques, single-mode waveguides that have non-linear optical response are challenging to manufacture and multimodeness is often introduced in the fabrication process by the chosen waveguide shape and size \cite{Bierlein1987, Fiorentino2007, Machulka2013}.

The multimodeness can also be taken advantage of, as is the case in Bragg-reflection waveguides (BRWs) that are built from semiconductor compounds \cite{Helmy2011}. The mode structure of BRWs is highly complex, and apart from supporting the total-internal-reflection modes they also carry higher-order modes, such as the Bragg modes \cite{Valles2013}. This is necessary for their use in frequency conversion processes. Recent demonstrations show that semiconductor waveguides are a versatile source for four-wave mixing \cite{Wathen2014} and parametric down-conversion \cite{Lanco2006, Horn2012}.

Here, we present an extension for the existing Fabry-Perot technique of linear optical characterization for multimode waveguides.  We apply this method to BRW, which has a very rich mode structure.
We resolve the modal characteristics of the waveguides with a single broadband measurement in the near infrared region (NIR). 
Similar to transmission spectroscopy in higher-order mode fibers \cite{Menashe2001, Nicholson2008}, our analysis uses the fact that different spatial modes generally possess different dispersion, and thus have different group indices.
From our results we extract a lower bound for the number of excited modes and compare their loss characteristics. Additionally, other properties such as the strength of the light coupling and the group effective index can be inferred for each mode. By investigating waveguides with different cross-sections we show that the multimodeness is increasing with a larger waveguide size.

BRWs are well-suited for loss measurements via the Fabry-Perot technique, due to the high refractive index ($>$3) of the semiconductor material. If their facets are uncoated, typically a Fresnel reflectivity of about 30\% is expected. Thus, the waveguide forms a low-finesse ($\mathcal{F}\approx 2.5$) Fabry-Perot resonator. The transmission spectrum of this resonator can be readily calculated by summation of the phasors at the facets. Following the notation in \cite{Hofstetter1997}, the transmittance of this resonator in terms of a vacuum wavenumber $\beta=2\pi/\lambda$, where $\lambda$ is the vacuum wavelength of the light, is given by
\begin{widetext}
\begin{equation}
\label{eq:transmittivity}
T(\beta)=\frac{(1-R)^2\exp(-2kL\beta)+4\sin(\phi)}{[1-R\exp(-2kL\beta)]^2+4R\exp(-2kL\beta)\sin^2(\phi+nL\beta)},
\end{equation}
\end{widetext}
where $L$ is the physical length of the resonator (i.e. waveguide length), $R$ and $\phi$ are the reflectivity and phase change at the facet, respectively. In the dispersive case, the frequency-dependent effective index of the mode $n=n(\beta)$ can be substituted. Linear losses are modeled through a complex effective index $\tilde{n}=n+ik$, where $k$ is the absorption index that can be related to the linear loss coefficient $\alpha$ via the vacuum wavelength $\lambda_0$ via $k = \frac{\alpha\lambda_0}{4\pi}$. The Fourier-transformed transmission spectrum shows distinct peaks with a quasi-double-exponential shape, centered approximately at integer multiples of the optical length $n_gL$ of the dispersive resonator, with $n_g$ being the group index of the mode. These higher harmonics can be explained by applying the Fourier phase shift theorem to the phasors that make up the transmission spectrum: the light propagating along the waveguide accumulates some phase. A phase shift in real space causes a translational shift in the Fourier space. Consequently, the individual facet reflections are separated in the Fourier domain, analogous to higher harmonics in acoustics with the amplitude at each reflection being preserved. Thus, the ratio of the amplitudes of subsequent peaks indicates the total loss $\widetilde R$ at each pass, which is given by
\begin{equation}
\label{eq:Rtilde}
\widetilde{R}=Re^{-\alpha L}.
\end{equation}
If the facet reflectivity $R$ and waveguide length $L$ is known, the linear loss coefficient $\alpha$ is calculated straightforwardly.

The linearity of the Fourier transform is a key feature for applying the Fabry-Perot analysis to multimode waveguides. Essentially, individual Fabry-Perot resonators are formed for the different modes with their different propagation indices.
This is justified if the sample is designed in a way that avoids interaction between the orthogonal eigenmodes \cite{Rossi2005}. To facilitate this constraint in our BRWs, the cross-section of the waveguides along the propagation direction remains constant and adjacent waveguides are isolated by separating them far enough so that evanescent coupling is avoided.
Usually the spacing of modes in the effective index scale is narrow, and therefore, highly complicated beating effects are recorded in the transmission spectrum. 
However, the distinct propagation indices lead to distinct optical lengths, and hence, it is possible to observe and measure each mode individually in Fourier space. 
In addition to this the relative excitation strength of each mode is also preserved. This is of particular interest for analyzing BRW waveguide designs, whose properties rely on higher order spatial modes.

Strictly speaking, the multimode inverse Fabry-Perot transform is an ill-posed inverse problem as the number of contributing modes is not known a-priori. Fortunately, Fourier transforming the multimode transmission spectrum results in a sparse representation, which enables us to do modally resolved measurements without knowing the proper inverse. Note that system-dependent additional effects, for example limited spectral resolution, may influence the Fourier transform. Thus, the achievable measurement accuracy will vary depending on the studied waveguide and its optical properties. Here, simulations  are very useful for understanding the implications of a particular effect on the measurements (see Supplement). Figure \ref{fig:th-spectrum} shows a simulation of the expected spectrum of a two-mode system representative of our sample. Despite having the same loss parameters, the slopes are different because only the \emph{ratio} of subsequent peak amplitudes is related to the total loss.

 In contrast to the more common fringe contrast measurement, Fourier analysis takes into account all the recorded information, not only minima and maxima. Thus, it is more robust against noise and other corruptions. Due to the uncertainty principle, measuring several dozens of fringes is required to get a high resolution mode spectrum in Fourier space. Usually, a lot of information can already be gained by visual inspection and interpretation, since reading the spectrum is easy and intuitive.

\begin{figure}[htbp]
\centering
\includegraphics[width=\linewidth]{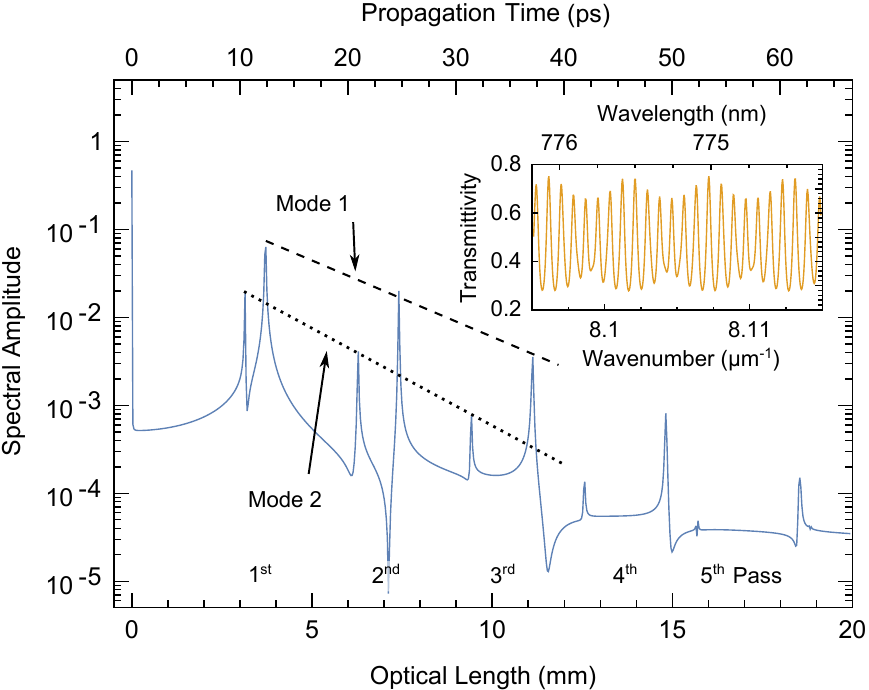}
\caption{
Simulated spectrum of the fringe pattern of two dispersive ($n_1=3.13$, $n_2=3.4$, with typical Al$_\mathrm{x}$Ga$_\mathrm{1-x}$As dispersion at 775 nm) and lossy modes ($R=0.3$, $k=10^{-5}$) representative of our 0.9 mm long sample. Excitation strength is 80\% for Mode 1 and 20\% for Mode 2. The peaks are expected to appear at multiples of the resonators optical lengths indicating the group effective index for each mode. The inset shows an excerpt of the simulated fringe pattern. There are five passes of each mode visible, each with their respective slope. }
\label{fig:th-spectrum}
\end{figure}

We are especially interested in the BRW properties in the near infrared region around 780 nm, since the Bragg mode in this wavelength regime is exploited to pump the process of parametric down-conversion \cite{Guenthner2014}. Our BRWs are fabricated with molecular beam epitaxy and have the layer structure described in \cite{Abolghasem2009_2, Horn2012}. The ridges are defined by electron beam lithography and transferred into the semiconductor via reactive plasma etching. We investigate BRW samples with two different lengths, 0.9 mm and 2.0 mm, which are cleaved pieces of a larger sample. Ridge widths from 3.0 to 6.5 $\mu$m in 0.5 $\mu$m steps were used in this work. Our BRWs are modally phase matched for nonlinear conversion processes between the NIR and telecom range. Contrary to the telecom range, the mode in the NIR is a higher order spatial mode.

\begin{figure}[htbp]
\centering
\includegraphics[width=\linewidth]{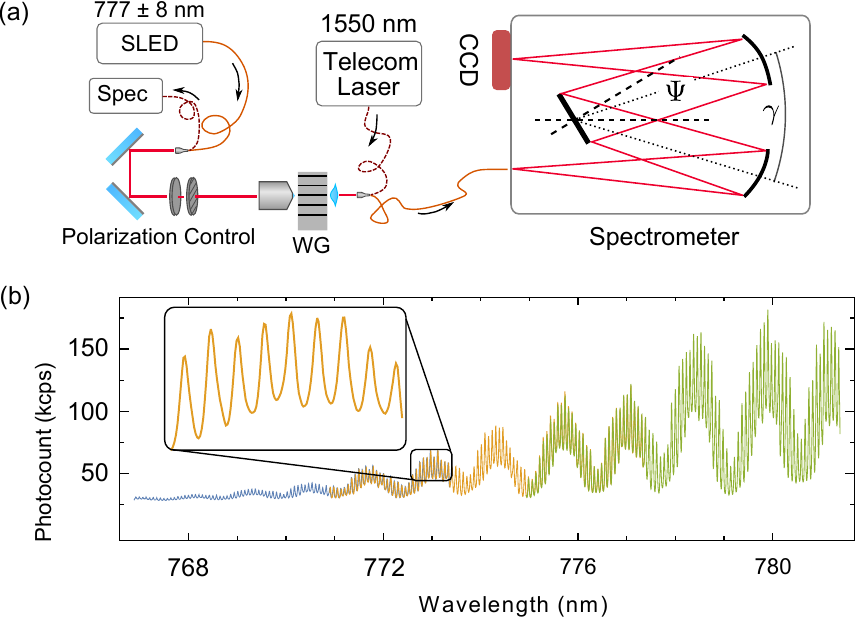}
\caption{(a) Experimental setup consisting of the superluminescent diode (SLED), polarization control optics (half-wave plate and sheet polarizer), coupling optics (microscope objective and aspheric lens) and the spectrograph. For optimization on SHG (see main text), a tunable telecom laser is backcoupled and analyzed with another spectrometer. $\Psi$ and $\gamma$ are the geometrical parameters of the spectrograph used for calibration (see Supplement). (b) Raw, 14.5 nm wide, stitched transmission spectrum. The different colors show the three exposures at different central wavelengths set at the spectrograph. The 210 Fabry-Perot fringes of the waveguide resonator can be seen superimposed on the oscillation (FSR $\approx$ 2 nm) of another resonator formed by additional optics in the beam path. The power envelope is determined by the spectrum of the SLED. }
\label{fig:ex-spectrum}
\end{figure}

\begin{figure*}[!htbp]
\centering
\includegraphics[width=\linewidth]{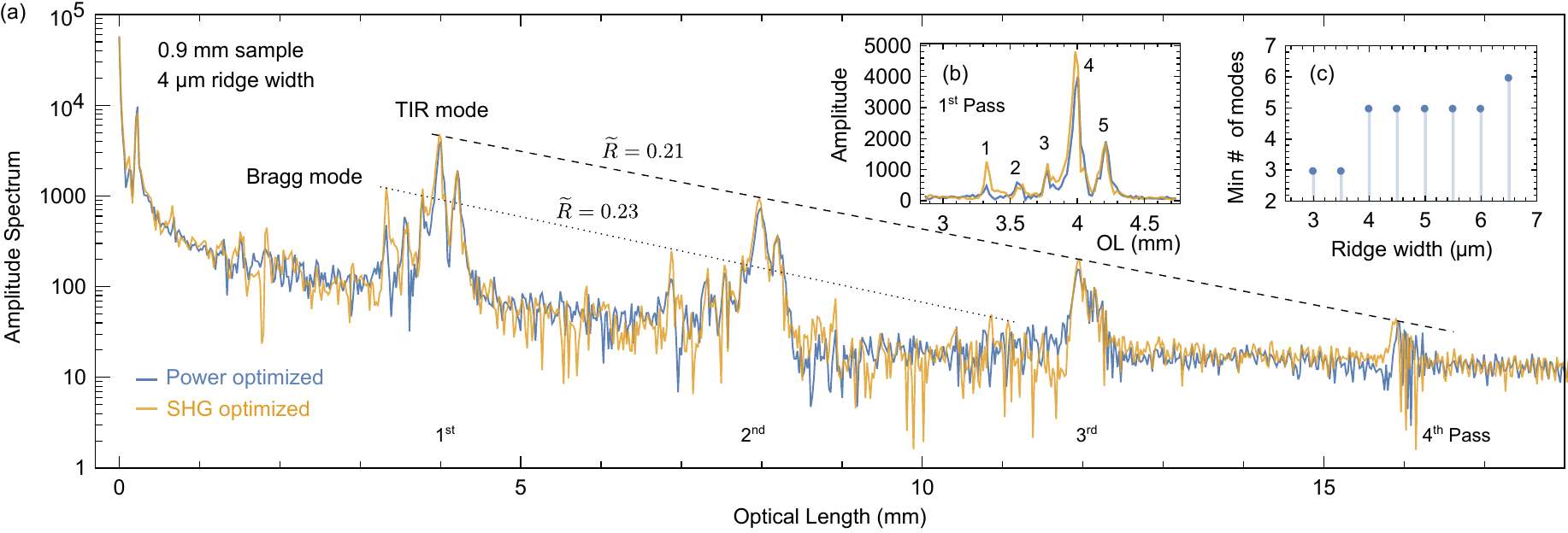}
\caption{(a) Modulus of the Fourier transform of the transmission spectrum from Fig. \ref{fig:ex-spectrum}(b) (blue solid line). Superimposed is a second spectrum showing the result from optimizing by maximizing SHG in the waveguide (orange solid line). The strongest mode is visible up to the $4^\mathrm{th}$ pass. Two modes and their respective total loss value $\widetilde{R}$ are indicated. (b) Linear plot of the first pass with five easily discernible modes. (c) The minimum visible mode numbers of several waveguides with different ridge widths. }
\label{fig:ex-shgopt}
\end{figure*}

In our experiment as illustrated in Fig. \ref{fig:ex-spectrum}(a), we launch the broadband light of a superluminescent diode (Superlum BLMS mini, centered at \SI{777}{\nano\meter}, \SI{17.4}{\nano\meter} FWHM) with power and polarization control into our BRW with a 100$\times$ microscope objective. The polarization of the input beam is kept horizontal. After the waveguide the transmitted light is collimated with an aspheric lens and coupled to a single-mode fiber. The single-mode fiber may cut out some modes or affect their coupling strengths, this is, however, a prerequisite for gaining the desired resolution. Nevertheless, in this configuration our measurement technique provides direct indication to what extent higher order modes can be coupled to or excited via a single mode fiber.
The collected light is then directed through an spectrograph incorporating a grating with a 1800 grooves/mm (Acton SP2750 with \SI{750}{\milli\meter} focal length), where we image the end facet of the single mode fiber onto a 2D camera with large sensor and small pixels ($9\times9$ $\mu$m$^2$, SBIG STT-1603ME). The achievable spectral resolution with this setup is better than \SI{10}{\pico\meter} FWHM across the entire image plane. Fig. \ref{fig:ex-spectrum}(b) shows a typical transmissions spectrum of our waveguides. In general, we employ two different procedures to optimize coupling into the Bragg mode: either maximizing the transmitted intensity or using a more complex scheme that involves second-harmonic generation (SHG). In the latter case, we couple a telecom laser backwards to generate the second-harmonic signal in the phase-matched Bragg-mode, which is then used to optimize the single-mode input coupling optics.

% SHG optimization

 Our results in Figure \ref{fig:ex-shgopt}(a) illustrate two different measurement configurations. First, we only optimize the transmission through the waveguide into the single mode fiber. In this case, shown with blue solid line in Fig. \ref{fig:ex-shgopt}(a), most of the input light couples to the fundamental total internal reflection (TIR) mode. After SHG optimization (yellow solid line in Fig. \ref{fig:ex-shgopt}(a)) we can easily identify the Bragg-mode and clearly notice that the amount of light coupled into the Bragg mode is largely increased, while the other peaks corresponding to other modes stay approximately the same. We note that, while most of the power is still coupled to other modes, the amplitude of the Bragg-mode is increased twofold. By integrating over the spectrum of the first pass, we estimate that in the SHG optimized case 8\% of the in-coupled power is guided by the Bragg-mode.
 Furthermore, in Fig. \ref{fig:ex-shgopt}(b) five excited modes can be identified for the given waveguide. As each mode has a different group index, we can separate the modes from the background by searching for a signal at twice the optical length. Fig. \ref{fig:ex-shgopt}(c) shows the number of clearly identifiable modes with respect to the ridge width. These values are only lower bounds on the minimum number of excited modes because closely spaced or degenerate modes cannot be separated. Nevertheless, a clear trend is visible that the number of modes increases with increasing ridge width.

 % Reflectivity

In order to determine the mode loss, Eq.~(\ref{eq:Rtilde}) has to be employed. Since only $\widetilde{R}$ and $L$ are known, we are left with two unknowns: the modal reflectivity $R$ and the loss value $\alpha$. We have two samples with different lengths (0.9 mm and 2.0 mm) but identical waveguides available. If we rewrite Eq.~(\ref{eq:Rtilde}) as a function of $L$, $\widetilde{R}(L)=Re^{-\alpha L}$ we can find a least-squares solution for $\alpha$ and $R$ simultaneously. In order to get good statistics, we measure multiple similar waveguides at each of the two waveguide lengths. Both values strongly depend on the exact value of $\widetilde{R}$, so corrections for the limited resolution have to be applied (see Supplement). For our BRWs, we retrieve a TIR mode reflectivity of $R = 0.35(4)$ and an ensemble loss coefficient of $\alpha=\SI{0.5(1)}{\per\milli\meter}=\SI{2.2(5)}{\dB/\milli\meter}$. The expected Fresnel reflectivity is 0.30 for an interface between a material with refractive index of 3.4 and air. However, the Fresnel equations tend to underestimate the semiconductor waveguide facet reflectivity. For systems similar to ours, we may say that the mode reflectivity is 20 \% higher than the Fresnel reflectivity \cite{Herzinger1993}. This is in good agreement with the measured value.
 For the modes of the waveguide in Figure \ref{fig:ex-shgopt}, taking the values of $\widetilde{R}$ and the facet reflectivity of $R = 0.35(4)$, yields linear loss coefficients of $\alpha_\mathrm{TIR}=\SI{0.46(12)}{\per\milli\meter}$ and $\alpha_\mathrm{Bragg}=\SI{0.36(12)}{\per\milli\meter}$. The uncertainty is mostly determined through the uncertainty of the facet reflectivity. A survey of many waveguides with different ridge widths indicates that total loss $\widetilde{R}$ varies only about 10\% for both modes. This suggests a high quality, highly repeatable waveguide fabrication and a lower level of loss than previously reported for BRW modes \cite{Bijlani2009, Horn2012}.

\begin{figure}[htbp]
\centering
\includegraphics[width=\linewidth]{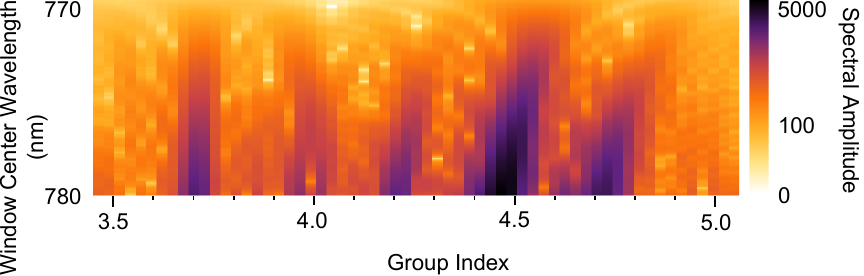}
\caption{Spectrogram of the first pass of the waveguide in Fig. \ref{fig:ex-shgopt}. In order to resolve the individual modes, a large fraction of the data ($\approx$ 70\%) was partitioned for each slice and weighted using a sinc window. }
\label{fig:spectrogram}
\end{figure}

  % Group effective index
 Finally, we estimate group velocities of $v_g = \SI{68(1)}{\micro\meter/\pico\second}$ and $v_g = \SI{81(1)}{\micro\meter/\pico\second}$ for the TIR and Bragg modes at the NIR. The group velocity is an important parameter for example when modelling the spectral properties of the twin photons created in the process of parametric downconversion, which for BRWs is typically gained from numerical simulations (see \cite{Guenthner2014} and references therein). Luckily, for highly dispersive resonators, the Fourier analysis provides direct experimental access to this parameter. Additionally, Fig. \ref{fig:spectrogram} shows the portion of the spectrogram of the first pass, which hints at the group velocity dispersion (GVD) of the modes.

% Summary
To summarize, we have shown how Fourier analysis of the optical transmission spectrum can provide useful information about the modal structure of multimode waveguides. We have applied this method to our BRW samples at NIR wavelengths, where their rich modal structure prevent the fringe contrast method from providing sufficient data.
From our results we can infer the linear characteristics in a mode resolved manner. Our technique, especially when combined with broadband spectroscopy, can be employed towards multimode waveguide characterization at many integrated optics platforms.

\bigskip \noindent ERC project \textit{EnSeNa} (257531); FWF project no. I-2065-N27; DFG Project no. SCHN1376/1-1.

\bigskip\noindent We thank M. Sassermann and Z. V\"{o}r\"{o}s for laboratory assistance and A. Sailer for technical support.

\bigskip \noindent See Supplementary Material for supporting content.

% Bibliography
%\bibliography{losses-abbv}
%\bibliography{losses}

%\newpage

%\documentclass[9pt,twocolumn,twoside]{optica-suppl-materials}
%\setboolean{shortarticle}{false}
%
%\usepackage{siunitx}
%
%\title{Modally Resolved Fabry-Perot Experiment with Semiconductor Waveguides: supplementary material}
%
%\author[1,*]{B. Pressl}
%\author[1]{T. G\"unthner}
%\author[1]{K. Laiho}
%\author[2]{J. Ge{\ss}ler}
%\author[2,3]{S. H\"ofling}
%\author[2]{C. Schneider}
%\author[2]{M. Kamp}
%\author[1,4]{G. Weihs}
%
%\affil[1]{Institut f\"ur Experimentalphysik, Universit\"at Innsbruck, Technikerstra{\ss}e 25, 6020 Innsbruck, Austria}
%\affil[2]{Technische Physik, Universit\"at W\"urzburg, Am Hubland, 97074 W\"urzburg, Germany}
%\affil[3]{School of Physics \& Astronomy, University of St Andrews, St Andrews, KY16 9SS, United Kingdom}
%
%\affil[*]{Corresponding author: Benedikt.Pressl@uibk.ac.at}
%
%\dates{Compiled \today}
%
%\doi{\url{http://dx.doi.org/10.1364/optica.99.099999.s1} [supplementary document doi]}
%
%\begin{abstract}
%This document provides supplementary information to “Modally Resolved Fabry-P\'{e}rot Experiment with Semiconductor Waveguides". We give details about the model used for simulating the transmission spectrum, discuss the issues of resolution and calibration when performing the measurements with a grating spectrometer and present a simplified scheme to correct for limited spectral resolution.
%\end{abstract}

%\setboolean{displaycopyright}{true}

%\begin{document}

%\maketitle
%\thispagestyle{fancy}
%\ifthenelse{\boolean{shortarticle}}{\abscontent}{}

\section*{Supplementary Material}
\textbf{This document provides supplementary information to \emph{Modally Resolved Fabry-Perot Experiment with Semiconductor Waveguides}. We give details about the model used for simulating the transmission spectrum, discuss the issues of resolution and calibration when performing the measurements with a grating spectrometer and present a simplified scheme to correct for limited spectral resolution.}

\section{Transmission spectrum: simulation and analysis}

The transmission spectrum of a simple, isolated ridge waveguide is determined by its reflectivity of the air interface and the phase shift accumulated along the propagation direction. In that sense, the waveguide forms a Fabry-P\'{e}rot resonator, with the transmitted light showing the interference of partial reflections accumulated through multiple round-trips. The transmission spectrum shows a series of distinct Lorentzian peaks, whose width is determined by the coefficient of finesse and, subsequently, through the reflectivity. The position of the peaks is determined by the single-pass phase 

\begin{equation}
\delta(\beta) = n(\beta) L \beta,
\end{equation}
where $L$ is the waveguide length, $\beta=2\pi/\lambda$ is the vacuum wavenumber of the mode and n($\beta$) is the dispersive effective index of the mode.

As motivated in the main text, assuming orthogonal modes \cite{Rossi2005_2} the total multimode spectrum can be calculated as the sum of multiple resonators' spectra as
\begin{equation}
\label{eq:sum}
I(\beta)=\sum_{n} x_i T_i(\beta),
\end{equation}
in which $x_i$ denotes the amount of light coupled to the $i$-th mode.

In Eq.~(\ref{eq:transmittivity}) , the normalized transmission spectrum of the $i$-th single Fabry-P\'{e}rot cavity is given by
\begin{equation}
\label{eq:transmittivity}
\begin{split}
&T_i(\beta)=\\
&\frac{(1-R_i)^2\exp(-2k_iL\beta)+4\sin(\phi)}{[1-R_i\exp(-2k_iL\beta)]^2+4R_i\exp(-2k_iL\beta)\sin^2(\phi+\delta_i(\beta))},
\end{split}
\end{equation}
 where $k$ is the absorption index and $R$ and $\phi$ are the facet reflectivity and the corresponding phase shift. The Fresnel equations provide good initial estimates of the reflectivity $R=\left[(n-1)^2+k^2\right]/\left[(n+1)^2+k^2\right]$ and phase shift $\phi=\arctan\left[-2k/(n^2+k^2-1)\right]$.  In the limit of very low loss $k \ll n$, however, the phase shift $\phi$ can be neglected. For a certain mode $i$, we assume only that the phase accumulated along the propagation direction $\delta_i(\beta)$ changes according to the effective index of the mode $n_i(\beta)$. Furthermode, we note that in the dispersive case, the spacing of the peaks in frequency domain, or free spectral range (FSR), is effectively given by the group index $n_g = n + \omega \frac{\partial n}{\partial \omega}$, with $\omega$ being the angular frequency of the light \cite{yariv2007photonics,siegman1986lasers}. This result carries on to the Fourier domain: the "optical length" measured from the peak position of each mode actually reports $n_g L$ in first approximation . The group velocity dispersion (GVD) may cause an additional shift of the peaks, for typical values, however, it is very small compared to the group index.

To simulate the measurement of the spectrometer, $I(\beta)$ is sampled in frequency space according to the pixel spacing of our spectrometer camera. The result is Gaussian filtered with a filter width equivalent to the measured resolution of the spectrometer (see Sec. \ref{sec:resolution}). Subsequent simulations show that the power variation of the utilized broadband light source (superluminescent diode) has little effect on the higher-order peaks appearing at multiple resonator passes, aside from slightly reducing the signal.

\section{\label{sec:resolution} Spectrograph considerations}

The FSR of the waveguide resonator scales inversely proportional to the group index and resonator length. Therefore, care has to be taken when using a spectrograph for recording the fringes, especially with regard to the resolution. The Lorentzian aspect of the peaks and the beating pattern has to be properly resolved.

 In our case, the group indices of the waveguide eigenmodes lie between 3.1 and 4.1, which leads to a free spectral range of only \SI{40}{\pico\meter} for a \SI{2}{\milli\meter} long sample. Also, in order to record more fringes, several exposures at different central wavelengths have to be stitched together.  A modified Czerny-Turner model \cite{IntelliCal2011} was used as calibration function, as the higher order polynomial model performed unsatisfactorily due to the large spacing of the lines in this wavelength range. The calibration model relates the observed wavelength $\lambda$ at the pixel position $\Delta x_\mathrm{cam}$ for a given "central wavelength" setting $\lambda_c$ of the spectrometer controller:
\begin{equation}
\label{eq:czerny}
\begin{split}
\lambda(\Psi, \Delta x_\mathrm{cam})&=\frac{d}{m}\left[\sin\left(\Psi-\frac{\gamma}{2}-\arctan\frac{\Delta x_\mathrm{in}}{f}\right)\right.\\
              &\left.+\sin\left(\Psi+\frac{\gamma}{2}+\arctan\frac{\Delta x_\mathrm{cam}}{f}\right)\right],\;\textrm{and}\\
\Psi(\lambda_c)&=\arcsin\left(\frac{m\lambda_c}{2d\cos\frac{\gamma}{2}}\right).
\end{split}
\end{equation}

\begin{figure}[!hbtp]
\centering
\includegraphics[width=\linewidth]{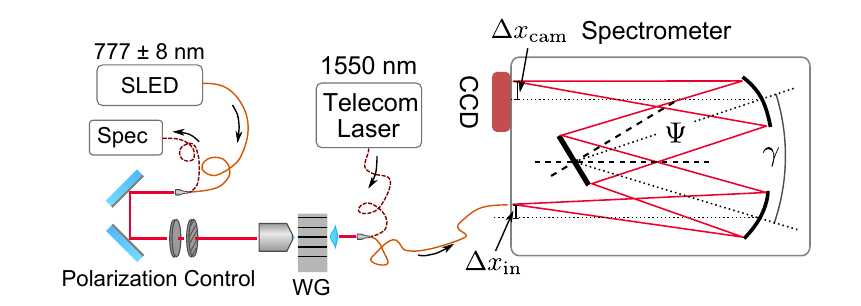}
\caption{Experimental setup for recording the transmission spectrum. Note that $\Delta x_\mathrm{in}$ is strongly exaggerated for illustration purposes.}
\label{fig:ex-setup}
\end{figure}

 Eq.~(\ref{eq:czerny}) is based on the key geometrical parameters of the spectrograph (see also Fig. \ref{fig:ex-setup}): $\gamma$ is the inclusion angle between grating and the mirrors, $\Psi$ is the grating angle calculated from $\lambda_c$, the diffraction order $m$ and the groove spacing $d$ of the grating. $\Delta x_\mathrm{in}$ and $\Delta x_\mathrm{cam}$ are the off-center distances of the input fiber and camera, respectively. During the measurement, only the central wavelength $\lambda_c$ was changed.

\begin{figure}[htbp]
\centering
\includegraphics[width=\linewidth]{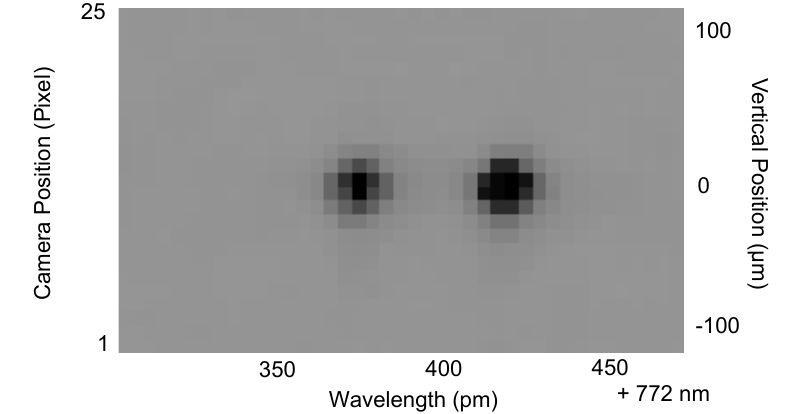}
\caption{Real space camera image of the Argon \SI{772.3761}{\nano\meter} and \SI{772.4207}{\nano\meter} lines.}
\label{fig:argon-lamp}
\end{figure}

 The calibration and resolution measurement was done by investigating several lines of a low-pressure Argon lamp between 766 and \SI{801}{\nano\meter} (see Fig. \ref{fig:argon-lamp}). These bright Argon lines served as reference for a total of 110 different ($\lambda_c$, $\Delta x_\mathrm{cam}$) pairs. Despite having few emission lines around \SI{775}{\nano\meter}, the typical calibration error is less than \SI{10}{\pico\meter} in the wavelength region of interest. This is more than enough to do a pixel-accurate stitch of three spectra with $\approx$ \SI{6.5}{\nano\meter} coverage each and an overlap of $\approx$ \SI{2.4}{\nano\meter}. This yields a total spectrum of \SI{14.5}{\nano\meter} showing approximately 210 (\SI{0.9}{\milli\meter} length sample, see main text Fig. \ref{fig:th-spectrum}) to 350 (\SI{2.0}{\milli\meter} length) fringes.

  % nonlinearity of calibration - vs. refractive index
 We note, that any nonlinearity on the wavelength calibration results in an artificial "dispersion", as there is an additional, frequency-dependent scaling error. The previously stated calibration error of \SI{10}{\pico\meter} over \SI{14.5}{\nano\meter} yields a nonlinearity of \SI{0.07}{\percent}. In comparison, the refractive index of the weakest dispersive material in our system, Al$_{0.8}$Ga$_{0.2}$As, changes approximately by \SI{0.2}{\percent} in the same wavelength range.

 The use of geometrical model of the spectrograph with the addition of $\Delta x_\mathrm{in}$ as extra parameter is critical in achieving the required calibration accuracy. However, more intricate geometrical models, for example with a ray optics approach, show that there is still some room for improvement \cite{Liu2013}.

 % Bild Auflösung

\section{Correcting for limited resolution}

The finite resolution of the spectrograph affects the values of the total loss $\widetilde{R}$, so here we show a simple correction scheme for this parameter. Recovery can be performed by recognising that limited resolution simply biases $\widetilde{R}$ towards lower values deterministically, depending on the group index of the mode and the resolution of the spectrometer. This is motivated by the properties of the Fourier transform: the recorded transmission spectrum is convolved with the point-spread function (PSF) of the spectrograph (see Fig. \ref{fig:argon-lamp}). For the sake of simplicity, we assume that the PSF can be represented as a narrow Gaussian whose width corresponds to the resolution. The individual pixels of the CCD can be seen a dirac comb which is convolved with the PSF to model the instrument response. In Fourier space, the convolution becomes a multiplication and the narrow Gaussian of the PSF transforms to a wide Gaussian with a width inversely proportional to the resolution. The multiplication of the Fourier transformed signal with the wide Gaussian causes damping at high optical lengths. This results in lower than expected values for $\widetilde{R}$. In that sense, the uncorrected loss values should be seen as upper bound.

\begin{figure}[!htbp]
\centering
\includegraphics[width=\linewidth]{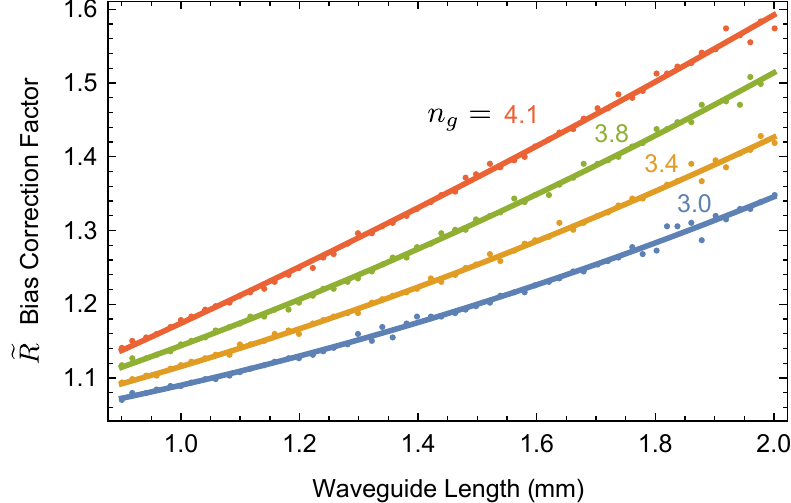}
\caption{Bias of the total loss $\widetilde{R}$ for several modes with different group indices for the resolution (10 pm) measured in our system. For example, with a 0.9 mm sample the measured values of $\widetilde{R}$ may be multiplied with a correction factor of $1.07-1.14$, depending on the mode.}
\label{fig:dispersion}
\end{figure}

Fig. \ref{fig:dispersion} shows correction curves for four modes with different group indices in the range expected in our sample. The corrected $\widetilde{R}$ is simply calculated by multiplying the measured value with the correction factor for a given waveguide length and group index. In the \SI{0.9}{\milli\meter} length sample from the main paper, the raw total loss values are expected to be overstated by about \SI{10}{\percent} on average. For the long, \SI{2.0}{\milli\meter} length samples, it may be as high as \SI{60}{\percent} in our setup.

\
 
A more elaborate method is deconvolving the original transmission spectrum, which is a well-studied technique in spectroscopy \cite{jansson1997deconvolution}. This approach has also been discussed in the context of the Fabry-Perot problem in microlaser gain measurements \cite{Guo2003}, but care has to be taken in order to preserve the signal-to-noise ratio and avoid the introduction of artifacts.

% Bibliography
%\bibliography{losses-abbv}

%\end{document} 

\end{document}